\documentclass[prx,
 reprint,
groupedaddress,
 aps,
floatfix
]{revtex4-2}

\usepackage{graphicx}
\usepackage{dcolumn}
\usepackage{bm}
\usepackage{siunitx} 
\usepackage{mathtools}
\usepackage{braket}
\usepackage{amsmath}
\usepackage{amssymb}	
\usepackage[colorlinks=true,linkcolor=blue,citecolor=blue,urlcolor=blue,bookmarksopen]{hyperref} 
\DeclareFontEncoding{LS1}{}{}
\DeclareFontSubstitution{LS1}{stix}{m}{n}

\usepackage{multirow}

\usepackage{tikz}
\definecolor{myblue}{rgb}{0.12156862745098039, 0.4666666666666667, 0.7058823529411765}
\definecolor{myorange}{rgb}{1.0, 0.4980392156862745, 0.054901960784313725}

\usepackage{hyperref}
\hypersetup{
    colorlinks=true,
    linkcolor=myblue,
    filecolor=magenta,      
    urlcolor=myblue,
    pdftitle={Optimal metrology with programmable quantum sensors},
    citecolor = myblue
}

\usepackage[english]{babel}
\usepackage{blindtext}

\newcommand{\cai}{$^{40}$\textrm{Ca}$^+$}
\newcommand{\MS}{\text{M\o{}lmer-S\o{}rensen}}
\usepackage{microtype}

\DeclareSIUnit{\dB}{dB}
\DeclareSIUnit{\belmilliwatt}{Bm}
\DeclareSIUnit{\dBm}{\deci\belmilliwatt}
\DeclareSIUnit{\belcarrier}{Bc}
\DeclareSIUnit{\dBc}{\deci\belcarrier}
\DeclareSIUnit{\beli}{Bi}
\DeclareSIUnit{\dBi}{\deci\beli}

\newcommand\Rx[1]{\def\arg{#1}\ifx\arg\empty \ensuremath{\mathcal{R}_x}\else \ensuremath{\mathcal{R}_x(#1)}\fi}
\newcommand\Ry[1]{\def\arg{#1}\ifx\arg\empty \ensuremath{\mathcal{R}_y}\else \ensuremath{\mathcal{R}_y(#1)}\fi}
\newcommand\Rz[1]{\def\arg{#1}\ifx\arg\empty \ensuremath{\mathcal{R}_z}\else \ensuremath{\mathcal{R}_z(#1)}\fi}

\newcommand\Tx[1]{\def\arg{#1}\ifx\arg\empty \ensuremath{\mathcal{T}_x}\else \ensuremath{\mathcal{T}_x(#1)}\fi}
\newcommand\Ty[1]{\def\arg{#1}\ifx\arg\empty \ensuremath{\mathcal{T}_y}\else \ensuremath{\mathcal{T}_y(#1)}\fi}
\newcommand\Tz[1]{\def\arg{#1}\ifx\arg\empty \ensuremath{\mathcal{T}_z}\else \ensuremath{\mathcal{T}_z(#1)}\fi}

\begin{document}

\title{Optimal metrology with programmable quantum sensors}

\author{Christian D. Marciniak$^{1}$}
\thanks{These authors contributed equally to this work}
\author{Thomas Feldker$^{1}$}
\thanks{These authors contributed equally to this work}
\author{Ivan Pogorelov$^1$}
\author{Raphael Kaubruegger$^{2,3}$}
\author{Denis V. Vasilyev$^{2,3}$}
\author{Rick van Bijnen$^{2,3}$}
\author{Philipp Schindler$^1$}
\author{Peter Zoller$^{2,3}$}
\author{Rainer Blatt$^{1, 2}$}
\author{Thomas Monz$^{1, 4}$}
\email{thomas.monz@uibk.ac.at}
\affiliation{$^1$ Institut f{\"u}r Experimentalphysik, 6020 Innsbruck, Austria}
\affiliation{$^2$ Institute for Quantum Optics and Quantum Information, 6020 Innsbruck, Austria}
\affiliation{$^3$ Center for Quantum Physics, 6020 Innsbruck, Austria}
\affiliation{$^4$ Alpine Quantum Technologies (AQT), 6020 Innsbruck, Austria}

\date{\today}
\begin{abstract}
Quantum sensors are an established technology that has created new opportunities for precision sensing across the breadth of science. Using entanglement for quantum-enhancement will allow us to construct the next generation of sensors that can approach the fundamental limits of precision allowed by quantum physics. However, determining how state-of-the-art sensing platforms may be used to converge to these ultimate limits is an outstanding challenge. In this work we merge concepts from the field of quantum information processing with metrology, and successfully implement experimentally a \emph{programmable quantum sensor} operating close to the fundamental limits imposed by the laws of quantum mechanics. We achieve this by using low-depth, parametrized quantum circuits implementing optimal input states and measurement operators for a sensing task on a trapped ion experiment. With 26 ions, we approach the fundamental sensing limit up to a factor of 1.45(1), outperforming conventional spin-squeezing with a factor of 1.87(3). Our approach reduces the number of averages to reach a given Allan deviation by a factor of 1.59(6) compared to traditional methods not employing entanglement-enabled protocols. We further perform on-device quantum-classical feedback optimization to `self-calibrate' the programmable quantum sensor with comparable performance. This ability illustrates that this next generation of quantum sensor can be employed without prior knowledge of the device or its noise environment.

\end{abstract}

\maketitle

\section{Introduction}

\begin{figure*}[ht]
    \centering
    \includegraphics[width=18cm]{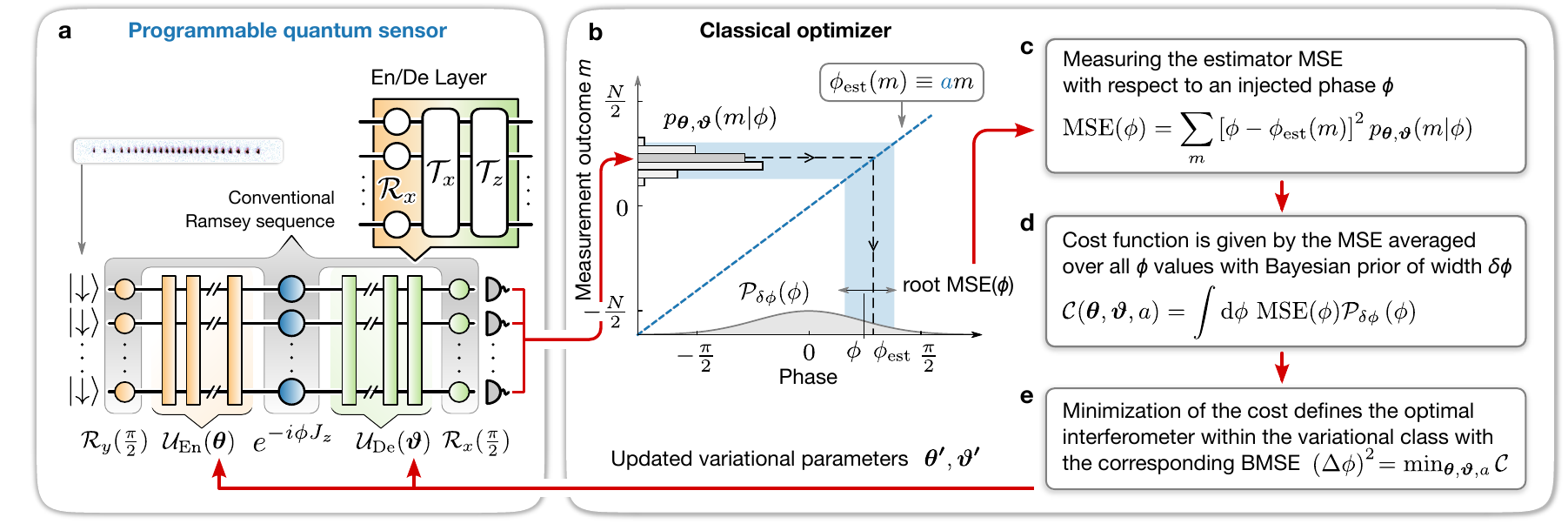}
    \caption{Measurement and feedback concept for variational quantum Ramsey interferometry circuits. 
    \textbf{a} A programmable quantum sensor executes a generalized Ramsey sequence with entangling and decoding unitaries $\mathcal{U}_\text{En}$, and $\mathcal{U}_\text{De}$ on $N$ particles. The unitaries are made from a repeating sequence of sensor resource gates, here collective qubit rotations $\mathcal{R}_{x, y, z}$, and infinite-range one axis twistings $\mathcal{T}_{x, y, z}$ with parameter sets $\left\{\bm{\theta}, \bm{\vartheta}\right\}$. 
    \textbf{b} Measurement of the collective spin $z$-projection results in a difference~$m$ of particles in $\ket{\uparrow}$ and $\ket{\downarrow}$ states. This is used to produce an estimate of the phase $\phi$ using a linear phase estimator $\phi_\text{est}$ with slope $a$. Prior knowledge of $\phi$ is encoded via the distribution $\mathcal{P}_{\delta\phi}$ taken here as Gaussian with variance $(\delta \phi)^2$ and zero mean.
    \textbf{c} The conditional probability $p_{\bm{\theta},\bm{\vartheta}}(m|\phi) = \big|\bra{m}\mathcal{R}_x(\tfrac{\pi}2)\,\mathcal{U}_{\rm De}(\boldsymbol{\vartheta})\, e^{-i\phi J_z}\, \mathcal{U}_{\rm En}(\boldsymbol{\theta})\mathcal{R}_y(\tfrac{\pi}2)\ket{\downarrow}^{\otimes N}\big|^2$ can be evaluated numerically or sampled in the experiment to calculate a mean squared error with respect to the true phase and the used estimator. 
    \textbf{d} An operational cost function $\mathcal{C}$ can be defined to quantify the interferometer performance for a variational sequence parameter set $\left\{\bm{\theta}, \bm{\vartheta}\right\}$. 
    \textbf{e} Minimization of the cost function is achieved by determining new parameter sets $\left\{\bm{\theta'}, \bm{\vartheta'}\right\}$ and comparing the associated costs either on-device or using classical simulation.}
    \label{fig:Overview}
\end{figure*}

Quantum sensing, that is using quantum systems to enable or enhance sensing, is arguably the most mature quantum technology to date. Quantum sensors have already found applications in many disciplines. The majority of these sensors are `quantum-enabled'; using the properties of a quantum system to perform a metrological task. Such applications have expanded rapidly including biology~\cite{taylor2016quantum,wu2016diamond}, medicine~\cite{rej2015hyperpolarized}, chemistry~\cite{frasco2009semiconductor}, or precision navigation~\cite{chen2019single} alongside traditional applications in physics such as inertial sensing~\cite{ahn2020ultrasensitive,moser2013ultrasensitive,chaste2012nanomechanical} or timekeeping~\cite{ludlow2015optical}. Quantum-enabled sensors perform close to or at the standard quantum limit (SQL) which originates from the quantum noise of the classical states used to initialise them. The latest generation of sensing technologies is going beyond the SQL by employing entangled states. These `quantum-enhanced' sensors are used in gravitational wave astronomy~\cite{tse2019quantum}, allow crossing the long-standing photodamage limit in life science microscopy~\cite{casacio2021quantum}, and promise improved atomic clocks~\cite{Vuletic2020}. However, these existing quantum-enhanced sensors, while beating the SQL, do not come close to what is ultimately allowed by quantum mechanics~\cite{gorecki2020pi}. Convergence to this ultimate bound is an open challenge in sensing~\cite{theory_posse}.

A parallel development in quantum technology that has seen massive progress alongside quantum sensing is quantum information processing, pursuing a `quantum advantage' in computation and simulation on near-term hardware~\cite{PreskillNISQ}. A crucial capability that has been developed in this context is the targeted creation of entangled many-body states~\cite{OmranGHZ, AQTION, Scholl2021Rydberg, Ebadi2021Rydberg, semeghini2021Rydberg}. A promising strategy there is to employ low-depth variational quantum circuits through hybrid quantum-classical algorithms~\cite{Peruzzo2014, Kandala2017, Kokail2019, cerezo2020variational}. Integrating this ability to program tailored entanglement into all aspects of sensing --- including measurement protocols~\cite{davis2016approaching, hosten2016quantum} --- will allow the construction of the next generation of sensors, able to closely approach fundamental sensing limits. The concept of such a `programmable quantum sensor' can be implemented on a great variety of hardware platforms, and is applicable to a wide range of sensing tasks. Moreover, their programmability makes such sensors amenable to on-device variational optimization of their performance, enabling an optimal usage of entanglement even on noisy and non-universal present-day quantum hardware.

Here we demonstrate the first experimental implementation of a programmable quantum sensor~\cite{kaubruegger2019variational} performing close to optimal with respect to the absolute quantum limit in sensing. We consider optimal quantum interferometery on trapped ions as a specific but highly pertinent example that promises applications ranging from improving atomic clocks and the global positioning system to magnetometry and inertial sensing. Our general approach is to define a cost function for the sensing task relative to which optimality is defined. We employ low-depth variational quantum circuits to search for and obtain optimal input states and measurement operators on the programmable sensor. This allows us to apply on-device quantum-classical feedback optimization, or automatic `self-calibration' of the device, achieving a performance close to the fundamental optimum.


\subsection*{Optimal quantum interferometry}
Our study below aims at optimal Ramsey interferometry to estimate a phase $\phi$ (Fig.~\ref{fig:Overview}~{\bf a}). In this context we aim to identify a suitable metrological cost function to quantify optimality. An established metric here is the mean squared error $\mathrm{MSE}(\phi)$ whose minimization yields the best \emph{average} signal-to-noise ratio for phase estimation at fixed signal. Traditionally, the optimization is done \emph{locally}, i.e.~for a small neighbourhood of phases around an a priori-specified value. This is achieved in the Fisher information approach, which underlies the discussion of Ramsey interferometry with squeezed spin states (SSS)~\cite{wineland1992spin} and, in particular, GHZ states~\cite{bollinger1996optimal}. Within this local approach the GHZ states are shown to saturate the so-called Heisenberg limit~(HL)~\cite{Pezze2018}.

In contrast, we are interested in an optimization for a \textit{finite} phase range $\delta\phi$, given by the desired dynamic range of the interferometer~\cite{gorecki2020pi, Degen2017}. This choice is motivated by applications using single-shot measurements such as in atomic clocks~\cite{Leroux_2017,theory_posse}.
We highlight that in frequency estimation applications the phase $\phi$ acquired during interrogation is not restricted to the $[-\pi,\pi)$ interval. Therefore the effect of the phase slipping outside this interval has to be taken into account as it leads to a permanent error in the frequency estimation.~\cite{Leroux_2017, macieszczak2014bayesian,theory_posse}. Under these circumstances the optimization may be accomplished in a Bayesian approach to optimal interferometry~\cite{macieszczak2014bayesian}, where a prior distribution of the phase, $P_{\delta \phi}(\phi)$ with width $\delta \phi$ defined as the standard deviation, is updated by the measurement to a posterior distribution with smaller width $\Delta \phi$. Consequently, we find as the metrological cost function $\mathcal{C}$ the \emph{Bayesian} MSE (BMSE), $\mathcal{C} \equiv \int{\mathrm{d}\phi\, \textrm{MSE}(\phi)\mathcal{P}_{\delta\phi}( \phi)}$~(see Fig.~\ref{fig:Overview}~{\bf b} - {\bf e}), that is the posterior mean squared error characterizing the phase probability distribution given the measurement outcome~$m$, and whose minimum we identify here with $(\Delta\phi)^2$. The optimal quantum interferometer (OQI) is thus obtained by minimization of the cost $\mathcal{C}$, that is the BMSE, over all entangled input states $\ket{\psi_{\mathrm{in}}}$, general measurements $\mathcal{M}$, and estimator functions $\phi_{\mathrm{est}}(m)$~\cite{macieszczak2014bayesian}. We emphasize that the OQI with large $\delta \phi$ will differ greatly from SSS or GHZ state-based interferometers, which optimize for local phase sensitivity~$\delta \phi \rightarrow 0$~\cite{macieszczak2014bayesian, theory_posse}.
 
Our goal below is to closely approach the OQI on programmable quantum sensors. We pursue a variational approach to optimal quantum metrology~\cite{theory_posse}, using a limited set of quantum operations available on a specific sensor platform. We consider a generalized Ramsey interferometer with an \emph{entangling} operation $\mathcal{U}_\mathrm{En}$ preparing an entangled state $\ket{\psi_{\mathrm{in}}}$ from the initial product state $\ket{\downarrow}^{\otimes N}$ of $N$ particles, and a \emph{decoding} operation $\mathcal{U}_\mathrm{De}$ transforming a typical observable, e.g.~$z$-projection of collective spin, into a general measurement~(Fig.~\ref{fig:Overview}~{\bf a} and~\ref{sec:VRI}). The variational approach consists of an ansatz, where both $\mathcal{U}_\mathrm{En}$ and $\mathcal{U}_\mathrm{De}$ are approximated by low-depth quantum circuits. These are built from `layers' of basic resource gates, which are given here by collective Rabi oscillations (qubit rotations) and collective entangling operations, commonly called infinite-range one axis twisting~(OAT) interactions~\cite{kitagawa1993squeezed}~(see~\ref{sec:RamseySequence} Eqs.~\ref{eq:rotations}) due to their action on the Bloch sphere. These resources are available in many atomic or trapped ion systems~\cite{Vuletic2020,bohnet2016quantum}. A quantum sensor is then programmed by specifying variational quantum circuits through $\mathcal{U}_\mathrm{En}(\bm{\theta})$ and $\mathcal{U}_\mathrm{De}(\bm{\vartheta})$, consisting of $n_\text{En}$ and $n_\text{De}$ `layers', respectively. These circuits define the conditional probability $p_{\bm{\theta},\bm{\vartheta}}(m|\phi)$, which describes the statistics of measurement outcomes $m$ given an input phase $\phi$. Together with a choice of phase estimator $\phi_{\mathrm{est}}(m)$ it determines the MSE, and in turn together with the prior $\mathcal{P}_{\delta\phi}$ the cost function $\mathcal{C}$. By varying the parameter vectors $\bm{\theta}$ and $\bm{\vartheta}$ we can therefore optimize the programmable quantum sensor for a given sensor platform and task. We refer to Methods~\ref{sec:VRI} for a technical summary and to Ref.~\cite{theory_posse} for details and intuitive explanation of the method (see II. C).

We implement the optimal Ramsey interferometry above on a compact trapped-ion quantum computing platform~\cite{AQTION}. This platform is used as a programmable quantum sensor, where in this work a linear chain of up to 26 \cai{} ions is hosted in a Paul trap. Optical qubits are encoded in the ground state $\ket{\textrm{4\,S}_{1/2},m_J = -1/2}$ and excited state $\ket{\textrm{3\,D}_{1/2},m_J = -1/2}$, which are connected via an electric quadrupole clock transition near \SI{729}{\nano\meter}. Technical details of the implementation can be found in the Supplementary Material, in particular state preparation and readout (S1), implementation and calibration of unitaries via the \MS{} interaction (S2), and technical restrictions imposed on the scheme (S3).

\begin{figure*}[htb]
    \centering
    \includegraphics[]{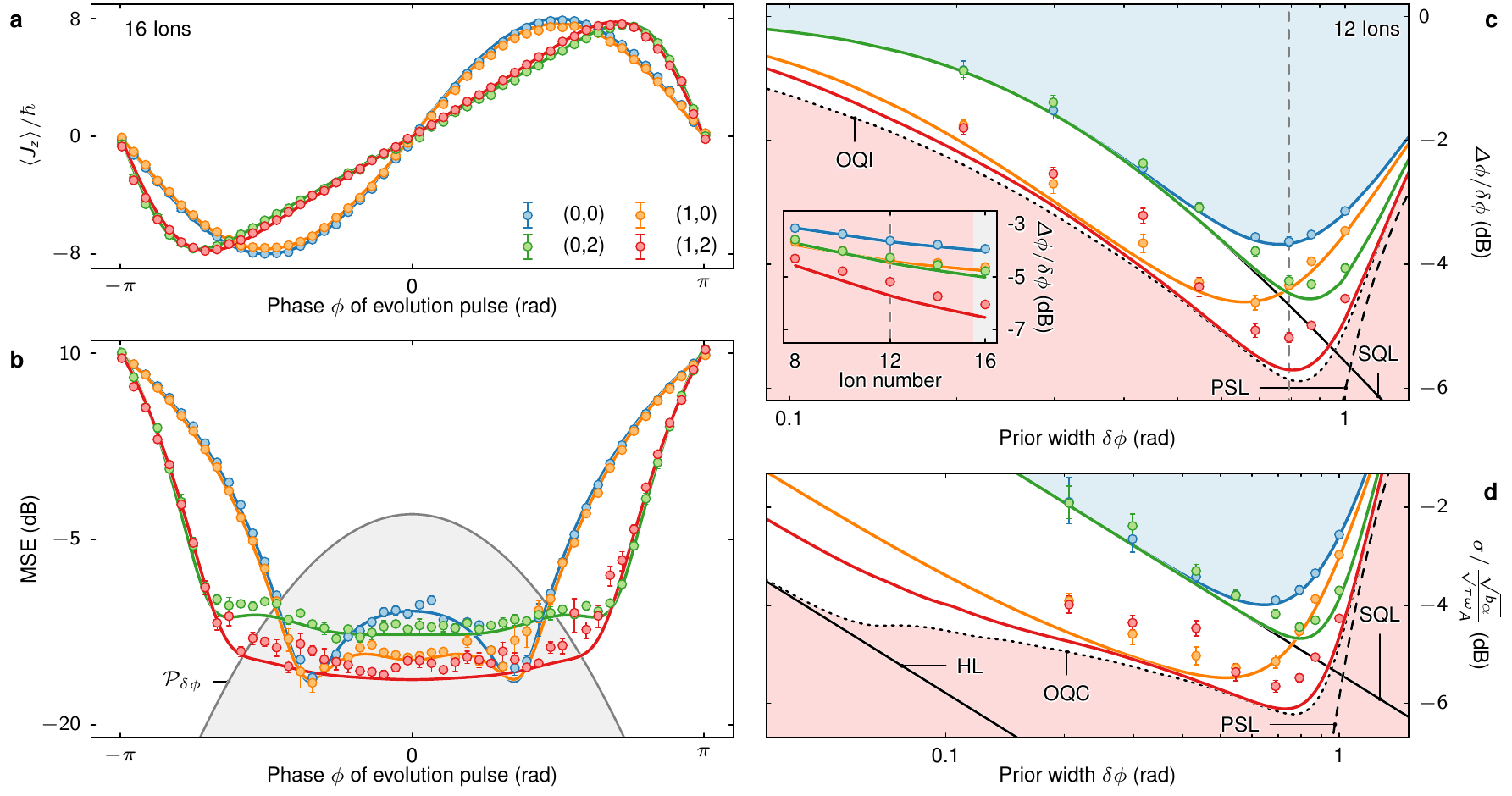}
    \caption{Generalized Ramsey sequence performance measurements. Markers are experimental data with 1$\sigma$ statistical uncertainties, and solid lines are theory with no free parameters. \textbf{a} 16 ion chain expectation value of spin operator $\left<J_z\right>/\hbar$ as a function of evaluation pulse phase $\phi$. 50 experiments are averaged per point, and 5 full traces are recorded to calculate mean and standard deviation of each curve. \textbf{b} MSE calculated from traces in \textbf{a} for optimal linear estimator $\phi_\text{est}$. Overlaid in grey is the prior distribution $\mathcal{P}_{\delta\phi}\left(\phi\right)$ with $\delta\phi \approx 0.79$ (minimum of BMSE vs $\delta\phi$).
    \textbf{c} $\Delta\phi/\delta\phi$ as a function of prior width $\delta\phi$ for 12 ions in the four variational sequences. The blue shaded region corresponds roughly to classical Ramsey sequences. The red shaded region is inaccessible to single-shot measurement schemes, where the boundary corresponds to the optimal quantum interferometer (OQI)~\cite{macieszczak2014bayesian}. Each point is produced by numerically integrating curves as in \textbf{b} over $\phi$ as per Eq.~\ref{eq:MSE}. The OQI, the standard quantum limit (SQL) and the phase slip limit (PSL) are indicated by black lines~(Methods~\ref{sec:Limits}). \textbf{Inset} $\Delta\phi/\delta\phi$ as a function of particle number at the prior width indicated by the dashed line. $N=16$ points calculated from data in \textbf{b}. \textbf{d} Allan deviation normalized to noise bandwidth $b_\alpha$, averaging time $\tau$ and reference frequency $\omega_A$ as a function of prior width with PSL, SQL, Heisenberg limit (HL)  and optimal quantum clock (OQC) based on the OQI are  indicated by black lines (Methods~\ref{sec:Allan}). Raw data, color scheme, and markers from \textbf{c}.}
    \label{fig:MSE_combined}
\end{figure*}


\section{Results}
We study the performance of the variationally optimized Ramsey sequences for four different choices of entangling and decoding layer depths $(n_\text{En}, n_\text{De})$ generating four distinct circuits: $(0, 0)$ being a classical coherent spin state (CSS) interferometer~\cite{Pezze2018} as the baseline comparison. All other sequences have been variationally optimized, $(1, 0)$ being similar to a squeezed spin state (SSS) interferometer~\cite{Pezze2018}, $(0, 2)$ with a CSS input state and tailored measurement, and finally $(1, 2)$ with both tailored input and measurement.

\subsection*{Direct implementation of theory parameters}
Following the execution of a Ramsey sequence (Eq.~\ref{eq:SequenceOriginal} in~\ref{sec:VRI}, or S2) we perform projective measurements at different Ramsey phases $\phi$ to reconstruct the expectation value of the total spin $z$ projection, $J_z$ (Fig.~\ref{fig:MSE_combined} \textbf{a}). From the measurements we construct the MSE (Fig.~\ref{fig:MSE_combined} \textbf{b}) using the linear estimator function $\phi_\text{est} = a m$ with slope $a$ which minimizes the cost function $\mathcal{C}$ obtained from integration according to Eq.~\ref{eq:cost} (see S4 and SI Tab.~1 for calculation, and S9 for discussion of other estimators). Qualitatively, Ramsey sequences with input state squeezing ($n_\text{En}>0$) dip below the CSS around $\phi=0$ as seen in Fig.~\ref{fig:MSE_combined} \textbf{b}. This dip is a manifestation of reduced projection noise. Sequences with optimized measurement operators ($n_\text{De}>0$) exhibit a broader range of $\phi$ values for which the MSE is comparable to the $\phi=0$ value. This is a consequence of the enhanced dynamic range that the non-trivial decoding unitaries impart, that is, the range over which the expectation value $\left<J_z\right>/\hbar$ remains well-approximated by the linear estimator (Fig.~\ref{fig:MSE_combined} \textbf{a}). Combining tailored input and measurements ($n_\text{En}, n_\text{De}>0$) yields an MSE which is both lower and wider than the CSS baseline.


To study this behaviour quantitatively as a function of the prior width $\delta\phi$ and particle number $N$ we calculate the BMSE scaled to the prior width $\delta\phi$ used. This is a convenient measure since $\delta\phi$ encapsulates \textit{prior} knowledge of $\phi$ and $\Delta\phi$ encapsulates \textit{posterior} knowledge after measurement. Their ratio $\Delta\phi/\delta\phi$ is therefore  bounded on the interval $[0, 1]$. We investigate this quantity for $\delta\phi\in\left[0.2, 1\right] \text{rad}$ as a representative sample of the parameter space, since no information is gained as $\delta\phi\rightarrow 0$, due to quantum projection noise overwhelming the signal, or $\delta\phi\rightarrow\pi$, due to phase slips outside the interval of unambiguous phase estimation~(Fig.~\ref{fig:MSE_combined}~\textbf{c}). For more details see Methods~\ref{sec:Limits}.

All variationally optimized sequences outperform the CSS within this measure (Fig.~\ref{fig:MSE_combined} \textbf{c}). The effect of change in dynamic range is evident in the location of a sequence's minimum. Minima of sequences with decoding layers shift towards larger prior widths with respect to the CSS, while for the direct spin-squeezing $(1, 0)$ it shifts towards smaller values. Sequences with a larger number of operations deviate more strongly from the theory predictions due to accumulation of gate errors. This behaviour is consistent across a range of particle numbers (Fig.~\ref{fig:MSE_combined} \textbf{c} inset). The deviation decreases as the system size does. We attribute this to the decrease in the fidelity of entangling operations~\cite{AQTION}.

The $(1, 2)$ scheme outperforms all others despite the increased complexity. In particular, it outperforms the simple spin-squeezing $(1, 0)$ scheme at both the optimal $\delta\phi$ for $(1, 2)$ and $(1, 0)$ approaching closely the OQI (see Tab.~\ref{tab:BMSEGains}). Specifically, for 26 particles and at their respective optimal prior widths, the $(1, 0)$ sequence approaches the OQI up to a factor of 1.87(3) (or \SI{2.73(7)}{\dB}), and the $(1, 2)$ sequence up to a factor of 1.45(1) (or \SI{1.61(2)}{\dB}). At this optimal prior width the $(1, 2)$ sequence would reduce the required number of averages to achieve the same Allan deviation as a classical Ramsey sequence by a factor of 1.59(6). A pictorial interpretation, in terms of Wigner distribution, of the optimized (optimal) interferometer can be found in Ref.~\cite{theory_posse}.

\begin{table}[h]
    \renewcommand\arraystretch{1.1}
    \centering
    \begin{tabular}{c|c|c||c|c}
          & \multicolumn{2}{c||}{$N=12$} & \multicolumn{2}{c}{$N=26$}   \\\hline
        $\delta\phi$ & 0.6893 & 0.792 & 0.5480 & 0.7403\\\hline
        $(0,0)$ & \SI{-3.56(3)}{\dB} & \SI{-3.63(8)}{\dB} & \SI{-3.22(3)}{\dB} & \SI{-4.53(3)}{\dB} \\
        $(1,0)$ & \SI{-4.61(12)}{\dB} & \SI{-4.34(4)}{\dB} & \SI{-5.63(7)}{\dB} & \SI{-5.39(2)}{\dB} \\
        $(1,2)$ & \SI{-5.06(11)}{\dB} & \SI{-5.18(8)}{\dB} & \SI{-5.84(9)}{\dB} & \SI{-6.75(2)}{\dB} \\\hline
        OQI & \multicolumn{2}{c||}{$\SI{-5.86}{\dB}$} & \multicolumn{2}{c}{$\SI{-8.36}{\dB}$}
    \end{tabular}
    \caption{Comparison of measured values of $\Delta\phi/\delta\phi$ at two values of $\delta\phi$ corresponding to the minima of the $(1, 0)$ (smaller $\delta\phi$) and $(1, 2)$ (larger $\delta\phi$) scheme, respectively. Note that the location of the minimum for $(1, 2)$ and $(0 ,0)$ is identical to within the measurement resolution presented. For reference, the minimum of the optimal quantum interferometer (OQI, border to shaded red region in Fig.~\ref{fig:MSE_combined} \textbf{c}) is given as well.}
    \label{tab:BMSEGains}
\end{table}

For atomic clock settings $\Delta\phi$ can be rescaled to calculate the Allan deviation of a deadtime-free clock (see~\ref{sec:Allan}), as shown in Fig.~\ref{fig:MSE_combined} \textbf{d} given the same raw data.

\subsection*{On-device quantum-classical feedback optimization}
\begin{figure*}[ht]
    \centering
    \includegraphics[]{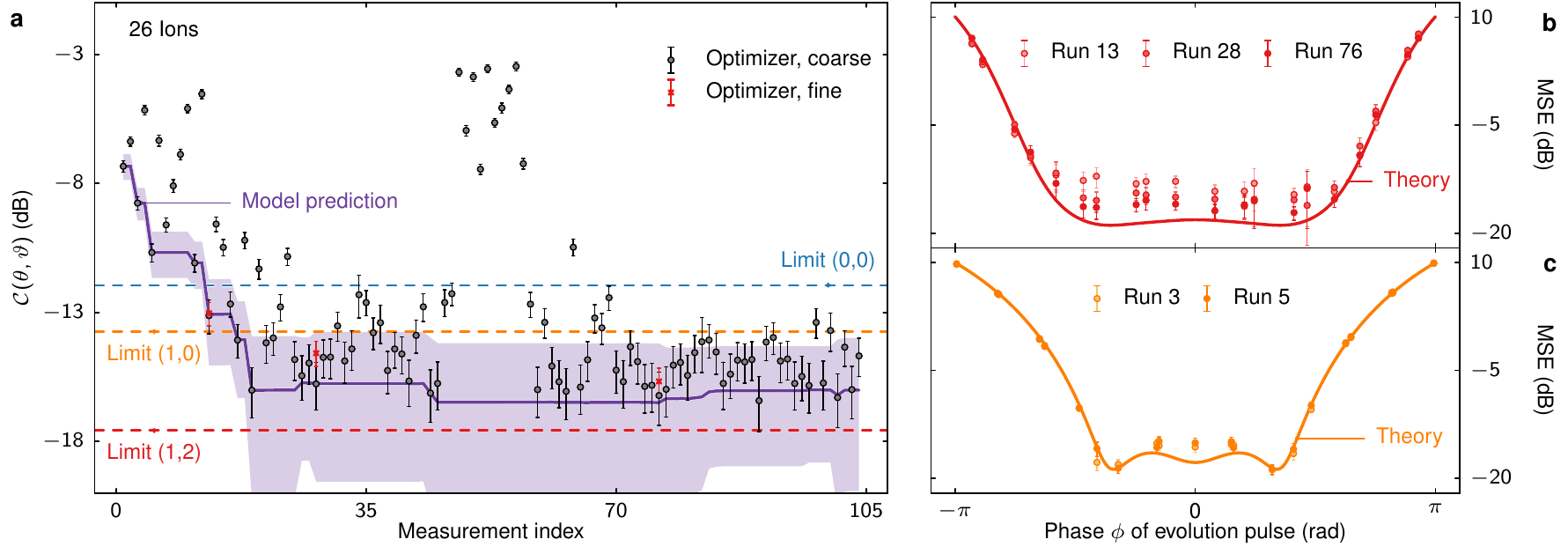}
    \caption{On-device hybrid quantum-classical optimization performance with 26 ions at $\delta\phi \approx 0.74$ (minimum BMSE vs $\delta\phi$), all error bars are 1$\sigma$ statistical uncertainties. \textbf{a} Optimizer cost function $\mathcal{C}$ as a function of measurement index (runs). Estimate is based on integration using 10 Hermite-Gauss nodes (S4) and 100 repetitions per point. Dashed lines indicate the achievable performance with the indicated sequences. Red crosses correspond to the automated fine scans displayed in panel \textbf{b}. \textbf{b} Automated fine scans (S) of the MSE with 20 nodes and 250 repetitions for three measurement indices. \textbf{c} Analogous fine scans from optimizer run using $(1, 0)$ sequence. Estimates based on 21 nodes and 250 repetitions per point.}
    \label{fig:Optimizer}
\end{figure*}

We further investigate the parameter `self-calibration' of the scheme in a regime where manual calibration is challenging, such that we expect direct application of theoretically optimal angles to no longer perform well. In particular, this is a regime where accurately calibrating the twisting parameters in $(\bm{\theta}, \bm{\vartheta})$ is no longer feasible. Minimization of the cost function is therefore achieved by a feedback loop where a classical optimization routine proposes new parameter sets to trial based on measurements performed on the quantum sensor. We employ a global, gradient-free optimization routine with an internal representation or `meta-model' of the cost function (S5).

The meta-model uses the known structure of the resource operations to learn an estimate of the cost function landscape based on the measurements, as seen in Fig.~\ref{fig:Optimizer} \textbf{a} for a 26 ion chain and the $(1, 2)$ circuit. Calibration of twisting angles is performed at a lower ion number (20), and then approximately scaled to the larger number. The cost function estimates are below the competing CSS $(0,0)$ and direct spin-squeezing $(1,0)$ after $\approx 20$ measurements despite this lack in accurate calibration. A full iteration of the algorithm is completed after $\approx50$ measurements in Fig.~\ref{fig:Optimizer} \textbf{a}.

Measurement points that the algorithm deems promising candidates for a minimum are resampled using `fine' scans (S6). Fine scans serve to increase the algorithm's confidence about predictions made on sparse data by better sampling, and relaxing symmetry assumptions of `coarse' scans. Fine scans show convergence towards the theory optimum as the algorithm progresses (Fig.~\ref{fig:Optimizer} \textbf{b}). Convergence is achieved more rapidly for the $(1, 0)$ sequence (Fig.~\ref{fig:Optimizer} \textbf{c}) due to the lower number of variational parameters, and consequently smaller parameter space. This convergence in both sequences despite the inability to accurately calibrate is a manifestation of the optimizer's ability to learn and correct for correlated gate (calibration) errors.
    

\subsection*{Frequency estimation}
All measurements up to this stage were taken by driving rotations \Ry{\phi} with resonant laser pulses as a consequence of our technical implementation (S2). This allows for deterministic mapping of the $\phi$ space, but in atomic clock experiments the phase $\phi$ would instead be imparted by the residual detuning of the drive from the atomic reference under the influence of noise. To gauge the performance of a clock we perform frequency estimation experiments. We calculate the variance of the frequency estimator from the known injected noise for a standard CSS interferometer, and the $(1, 2)$ interferometer optimized for a prior width $\delta\phi \approx 0.69$ (S7).

\begin{figure}[ht]
    \centering
    \includegraphics[]{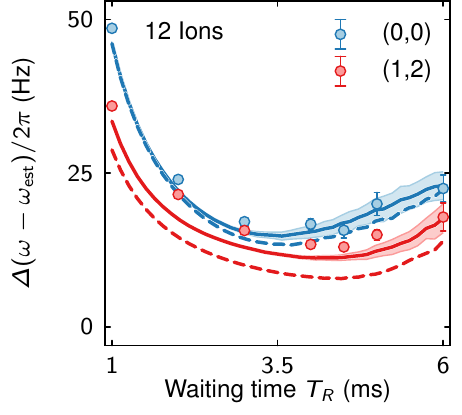}
    \caption{Frequency measurement using 12 ions with a standard and variationally optimized Ramsey sequence. Shown is the standard deviation of the difference between known injected and estimated frequency detuning. Markers are experimental data with uncertainty (finite sampling and projection noise). Dashed lines are theory simulations with no free parameters, while solid lines are simulation of measurement performance including a residual laser flicker noise of $b_\alpha\approx2\pi\cdot\SI{6}{Hz}$, and shading indicates error margin of simulation.}
    \label{fig:fig_clock}
\end{figure}

The optimized sequence outperforms the CSS for all considered Ramsey times (Fig.~\ref{fig:fig_clock}). In particular, this demonstrates robustness of the scheme with respect to variations in the prior width (Ramsey time, S8). The deviation between experiment and theory predictions can be explained by two observations. First, we independently measured predominantly frequency flicker noise of bandwidth $b_\alpha\approx2\pi\cdot\SI{6}{Hz}$ (S7) on the laser which is not present in the simple simulation. Second, the MSE used in the simulation is the ideal, theoretically-achievable one, while the experiment has deviations from the theory such as in Fig.~\ref{fig:MSE_combined} \textbf{b}. Simulating the metrology experiments with these additional noise sources restores good match between data and prediction. We note that this problem is not apparent in the BMSE or the Allan variance plots (Fig.~\ref{fig:MSE_combined} \textbf{c} and \textbf{d}) since it arises solely in the \Rz{} operation we employ here, while $\phi$ was imparted via \Ry{} there.

\section{Discussion and outlook}
Intermediate-scale quantum devices, acting as quantum sensors, provide the toolset to program entanglement and collective measurements to approach the ultimate limits of parameter estimation compatible with the laws of quantum physics. The present work has demonstrated programming a close-to-optimal quantum interferometer with (up to) $N=26$ entangled atoms on a trapped-ion quantum computer. A key element of our work has been to identify a pathway towards optimal quantum sensing by formulating it as a variational quantum algorithm, where circuits of increasing depths allow convergence towards the ultimate sensing limit. This limit is approached by optimizing the circuits using a task-specific cost function. The shallow quantum circuits used here are built from native, imperfect trapped-ion quantum operations, and are already shown to yield results close to optimal metrology. In a broader context, this suggests that the variational approach to optimal quantum sensing is both flexible and hardware efficient. When combined with the linearly-growing solution Hilbert space this indicates the potential to scale to significantly larger particle numbers.

Variational optimal metrology is immediately applicable to a wide range of sensing \textit{tasks}. Our demonstration of generalized Ramsey interferometry is relevant for atomic clocks, and we discuss the projected gains in Allan deviation in the Supplementary Material (S10). Furthermore, the prevalence of Ramsey interferometry in metrology renders our approach relevant to the measurement of magnetic fields~\cite{jones2009magnetic}, inertia~\cite{borde2002atomic}, displacement and electric fields~\cite{gilmore2021quantum}, as well as force measurements~\cite{gilmore2017amplitude}. While here we demonstrated quantum estimation of a single parameter, the present technique of variational optimal metrology readily generalizes to the multi-parameter case~\cite{Demkowicz_2020}.

Our approach is furthermore immediately applicable to other sensing \textit{platforms}. Variational quantum metrology can for example be implemented on programmable quantum simulators~\cite{kaubruegger2019variational} with well-established capabilities, in particular in higher spatial dimensions. While these readily scale to large particle numbers, they provide only non-universal entanglement operations via finite-range interactions. 
The `on-device' optimization of the metrological cost function, as demonstrated in the present work, then not only serves to find optimal input states and measurement protocols in presence of `real world' device imperfections and noise. Instead, it then also addresses the underlying computationally hard problem of preparation and manipulation of many-body quantum states. For increasing particle numbers this provides an example of a quantum device operating in a regime of \textit{relevant quantum advantage}, where many-body quantum states are both prepared and subsequently exploited in optimal metrology.



\newpage

\appendix
\renewcommand{\thesubsection}{M\arabic{subsection}}
\section*{Methods}

\subsection{Variational Ramsey interferometer} \label{sec:VRI} In variational Ramsey interferometry, the quantum sensor is initially prepared in the collective spin down state, and subsequently executes the variational Ramsey sequence given by
\begin{equation}
    \mathcal{U}_\text{R}(\phi, \bm{\theta}, \bm{\vartheta}) = \Rx{\tfrac{\pi}{2}}\mathcal{U}_\text{De}(\bm{\vartheta})\Rz{\phi}\mathcal{U}_\text{En}(\bm{\theta})\Ry{\tfrac{\pi}{2}},
    \label{eq:SequenceOriginal}
\end{equation}
where $\mathcal{R}_{x,y,z}$ are collective Rabi oscillations, and $\mathcal{U}_\text{En}(\bm{\theta}), \mathcal{U}_\text{De}(\bm{\vartheta})$ are entangling and decoding circuits, with control parameters $\bm{\theta}, \bm{\vartheta}$ (see Methods~\ref{sec:RamseySequence}). In between the two operations, the sensor interacts with an external field which imprints a phase $\phi$ onto the constituent particles. It is important to note that the Ramsey sequence is $2\pi$-periodic in $\phi$ and hence phases can only be distinguished modulo $2\pi$. 

After executing $\mathcal{U}_\text{R}(\phi, \bm{\theta}, \bm{\vartheta})$, we perform projective measurements of the collective spin yielding outcomes $m$ (difference of particles in $\ket{\uparrow}$ and $\ket{\downarrow}$). The phase $\phi$ is estimated from $m$ by means of a linear phase estimator $\phi_{\rm est}(m) = a\,m$,  which is optimal for the variational interferometer~\cite{theory_posse} and near-optimal for CSS- and SSS interferometers at the particle numbers considered here~\cite{andre2004stability}. We provide a quantitative comparison between the different estimation functions in S9.

The goal is to find parameters $\bm{\theta}, \bm{\vartheta}, a$, that give the best possible performance of the sensor. The performance of the sensor intended to correctly measure a given phase $\phi$ can be quantified by the mean squared error
\begin{equation}
     \textrm{MSE}(\phi) = \sum_m{\left[\phi-\phi_{\rm est}(m)\right]^2p_{\bm{\theta},\bm{\vartheta}}(m|\phi)},
     \label{eq:MSE}
\end{equation}
where 
\begin{equation}
    p_{\bm{\theta},\bm{\vartheta}}(m|\phi) = \big|\bra{m}\mathcal{U}_\text{R}(\phi, \bm{\theta}, \bm{\vartheta})\ket{\downarrow}^{\otimes N}\big|^2
\end{equation}
is the probability to observe a measurement outcome $m$, given $\phi$, and for given circuit parameters $\bm{\theta}, \bm{\vartheta}$.

In Bayesian phase estimation we are interested in a sensor that performs well for a range of phases $\phi$, occurring according to a prior distribution $P_{\delta \phi}(\phi)$. We assume $P_{\delta \phi}(\phi)$ to be a normal distribution with variance $(\delta \phi)^2$ and zero mean throughout, which is a choice particularly relevant for applications like atomic clocks. Note that the normal distribution has a finite probability for phase slips outside the unambiguous phase interval, determined by the period of $\mathcal{U}_\text{R}$. Phase slips contribute to the MSE, and dominate for $\delta \phi \gtrsim 1$~(see \ref{sec:Limits}).

A meaningful cost function for the sensor's overall performance is the average MSE, weighted according to the prior phase distribution $P_{\delta \phi}$ \begin{equation}
     \mathcal{C}(\bm{\theta}, \bm{\vartheta}, a)  = \int{\text{d}\phi\ \textrm{MSE}(\phi, \bm{\theta}, \bm{\vartheta})\mathcal{P}_{\delta\phi}\left( \phi\right)},
     \label{eq:cost}
\end{equation}
called Bayesian mean squared error (BMSE). 

In this work, the parameters $(\bm{\theta}, \bm{\vartheta}, a)$ are optimized with respect to the cost function $\mathcal{C}$, either numerically (see Fig.~\ref{fig:MSE_combined}), or on-device in a variational feedback loop (see Fig.~\ref{fig:Optimizer}). For on-device optimization, the parameter $a$ is held fixed at the numerically calculated optimal value.

To evaluate the cost function $\mathcal{C}$ in the variational feedback loop we run the Ramsey sequence, while exposing the sensor to a sequence of known \textit{injected} phases $\phi_i$. The cost function value is then estimated as
\begin{equation}
\mathcal{C}(\bm{\theta}, \bm{\vartheta}, a) \simeq \sum_i \textrm{MSE}(\phi_i) P_{\delta \phi}(\phi_i) w_i,
\end{equation}
where $w_i$ are Hermite Gaussian integration weights (see S4).

The minimum of the cost function $(\Delta \phi)^2 = \min_{\bm{\theta}, \bm{\vartheta}, a} \mathcal{C}$ can be interpreted \cite{Demkowicz2015} as 
\begin{equation}
    (\Delta \phi)^2 \approx \sum_m (\Delta \phi_m)^2 p(m)
\end{equation}
i.e. the variances$(\Delta \phi_m)^2$ of the posterior distributions $p(\phi|m)$ averaged according to the probability $p(m)$ to observe the measurement outcome $m$. Therefore we refer to $\Delta \phi$ as the posterior width. 

\subsection{Ramsey sequence}\label{sec:RamseySequence}
Following~\cite{theory_posse} the explicit form of the entangling and decoding unitaries are 
\begin{align}
    \mathcal{U}_\text{En} & = \prod_{k=1}^{n_\text{En}}\Rx{\theta_{k}^3}\Tx{\theta_{k}^2}\Tz{\theta_{k}^1}
    \label{eq:unitaries1}\\
    \mathcal{U}_\text{De} & = \prod_{k=1}^{n_\text{De}}\Tz{\vartheta_{k}^1}\Tx{\vartheta_{k}^2}\Rx{\vartheta_{k}^3}\label{eq:unitaries2}.
\end{align}
Here $\mathcal{R}_{x,y,z}$ are collective Rabi oscillations, and $\mathcal{T}_{x, y, z}$ are one-axis twisting (OAT) operations. Mathematically these operations can be represented as 
\begin{equation}
     \mathcal{R}_{x,y,z}(\beta) = \text{e}^{-i\beta J_{x, y, z}}\label{eq:rotations}, \quad
     \mathcal{T}_{x,y,z}(\chi) = \text{e}^{-i\chi J_{x, y, z}^2}
\end{equation}
where $\beta$ and $\chi$ are angles that depends on the interaction strength and time, and $J_{x, y, z}$ are collective spin operators in the Cartesian basis. We denote the collection of the three operations in Eqs.~\ref{eq:unitaries1}, \ref{eq:unitaries2} with the same subscript as one layer, and we denote by $n_\text{En}$ and $n_\text{De}$ the number of entangling and decoding layers. 


\subsection{Effects of resource restrictions}
\label{sec:Restrictions}
The globally optimal variational parameter sets depend on the ion number via the prior width. However, we may additionally restrict them based on platform constraints of fundamental or practical nature to find sets optimal with respect to device capabilities. This adds to the adaptability inherent to the scheme: We tailor the cost function to the sensing task, and the sequence resources, while parameter ranges are constrained by the experimental hardware. Combined this assists with assessing and interpretation of attainable results given real-world constraints. 

Furthermore, in systems of moderate size of order 50 and above this leads to the Allan deviation scaling down with particle number $N$ or Ramsey time $T_R$ at close to the ($\pi$-corrected) Heisenberg limit~\cite{gorecki2020pi,Chabuda:2020tv} up to a logarithmic correction~\cite{theory_posse, borregaard2013near}. This is of great practical utility in situations where measurements are made with a fixed budget in particle number or measurement time.

\subsection{Bounds on the Bayesian mean squared error}
\label{sec:Limits}
In the Bayesian framework, a bound on the BMSE in the limit of a narrow prior, $\delta \phi \ll 1$, is imposed by  quantum measurement fluctuations as captured by van Trees' inequality~\cite{Trees},
\begin{align}
(\Delta\phi)^{2} \geq  \frac{1}{ \overline{F}_{\phi} + \mathcal{I}}.
\end{align}
Here, the first term in the denominator is the Fisher information of the conditional probability, $\smash{F_{\phi}=\sum_{m}\big[\partial_{\phi}\log p(m|\phi)\big]^{2}p(m|\phi)}$, averaged over the prior distribution, $\overline{F}_{\phi}=\int d\phi\mathcal{P}(\phi) F_{\phi}$. The second term is the Fisher information
of the prior distribution, $\mathcal{I}=\int d\phi\mathcal{P}(\phi)\big[\partial_{\phi}\log\mathcal{P}(\phi)\big]^{2}$, representing the prior knowledge. 

For pure states of $N$ spin-$1/2$ particles, i.e. in the absence of decoherence, the Fisher information is limited by $F_{\phi}\le N^2$ which defines the Heisenberg limit~(HL)~\cite{Pezze2018}. In the case of uncorrelated states of atoms the Fisher information limit reads $F_{\phi}\le N$ and corresponds to the standard quantum limit~(SQL)~\cite{Pezze2018}. This results in SQL and HL limits on the BMSE, which read, respectively,
\begin{align}
\label{eq:SQL_BMSE}
    (\Delta \phi_{\rm SQL})^2 &= [N + (\delta\phi)^{-2}]^{-1},\\
    \label{eq:HL_BMSE}
    (\Delta \phi_{\rm HL})^2 &= [N^2 + (\delta\phi)^{-2}]^{-1}.
\end{align}
Here we used the Fisher information of a normal distribution with variance $(\delta\phi)^2$ for the prior, thus $\mathcal{I}=(\delta\phi)^{-2}$. Equations~\eqref{eq:SQL_BMSE},~\eqref{eq:HL_BMSE} define the corresponding limits in~Fig.~\ref{fig:MSE_combined}~{\bf c},~{\bf d} of the main text.

One can similarly define the $\pi$-corrected Heisenberg limit~\cite{gorecki2020pi} for the BMSE. This fundamental limit, however, is a tight lower bound only asymptotically in the number of atoms $N$. It becomes applicable for particle numbers, $N\gtrsim 100$, far beyond the size of our present experiment. Further details can be found in Ref.~\cite{theory_posse}.

A different kind of bound on the BMSE arises in the limit of large prior widths, $\delta\phi\gtrsim 1$, which we denote as the phase slip limit (PSL). The PSL is caused by phase slipping outside the interval of unambiguous phase estimation due to tails of prior distribution extending beyond the phase interval $[-\pi, \pi)$. We model the PSL as
\begin{equation}
    (\Delta \phi_{\rm PSL})^2 = \left[\left((2 \pi)^2 \times 2\int_{\pi}^{\infty}d\phi\, \mathcal{P}_{\delta\phi}(\phi)\right)^{-1} + (\delta\phi)^{-2}\right]^{-1}, 
\end{equation}
which is composed of the probability of phase slipping outside the $[-\pi, \pi)$ interval multiplied by the minimum squared error of $(2\pi)^2$ associated with the slip. The PSL gives rise to the increase of $\Delta\phi/\delta\phi$ values at $\delta\phi\gtrsim 1$ in~Fig.~\ref{fig:MSE_combined}~{\bf c},~{\bf d} of the main text.


\subsection{Allan deviation}
\label{sec:Allan}

In atomic clock settings the (Gaussian) prior distribution width $\delta\phi$ can be related to experimental system parameters, specifically the width of the distribution of expected phases after a Ramsey interrogation time $T_R$ subject to a noisy reference laser~\cite{theory_posse}. For a noise power spectral density $S(f) \propto f^{1-\alpha}$ of bandwidth $b_\alpha$ the functional form is given by
\begin{equation}
    \delta\phi = \left(b_\alpha T_R\right)^{\alpha/2}.    
\end{equation}

Based on this we can link the BMSE $\left(\Delta\phi\right)^2$ to  the Allan deviation as an established figure of merit in frequency metrology. For clock operation without deadtime, and with averaging time $\tau$ the Allan deviation $\sigma(\tau)$ is given by 
\begin{align}
    \sigma(\tau) &= \frac{1}{\omega_A}\frac{\Delta\phi_M}{T_R}\sqrt{\frac{T_R}{\tau}} = \frac{1}{\omega_A}\frac{\Delta\phi_M}{T_R\sqrt{n}}\label{eq:allan}\\
    \Delta\phi_M &= \Delta\phi/\sqrt{1-\left(\frac{\Delta\phi}{\delta\phi}\right)^2},
\end{align}
where $\Delta\phi_M$ is the effective measurement uncertainty of one cycle of clock operation~\cite{leroux2017line}. Here $n= \tau / T_{\rm R}$ is the number of measurements per averaging time, and $\omega_A$ is the (atomic) reference frequency. For a variational Ramsey sequence without decoder, i.e. $n_{\rm De} = 0 $ and in the limit of small $\delta\phi$, $\Delta\phi_M$ is determined by the Wineland squeezing parameter~\cite{wineland1994squeezed}  $\xi_W$, i.e. $\Delta\phi_M \rightarrow \xi_W/\sqrt{N}$. The equality holds as long as the Allan deviation is dominated by projection noise, and will break down once the contribution from laser coherence becomes appreciable.


\subsection*{Data availability}
All data obtained in the study is available from the corresponding author upon request.

\subsection*{Acknowledgements}
We gratefully acknowledge funding from the EU H2020-FETFLAG-2018-03 under Grant Agreement no. 820495. We also acknowledge support by the Austrian Science Fund (FWF), through the SFB BeyondC (FWF Project No.\ F7109), and the IQI GmbH.  This project has received funding from the European Union’s Horizon 2020 research and innovation programme under the Marie Skłodowska-Curie grant agreement No 840450. P.S. acknowledges support from the Austrian Research Promotion Agency (FFG) contract 872766.  P.S., T.M. and R.B. acknowledge funding by the Office of the Director of National Intelligence (ODNI), Intelligence Advanced Research Projects Activity (IARPA), via US ARO grant no. W911NF-16-1-0070 and W911NF-20-1-0007, and the US Air Force Oﬃce of Scientiﬁc Re-search (AFOSR) via IOE Grant No. FA9550-19-1-7044 LASCEM.

R.K., D.V.V., and P.Z. are supported by the US Air Force Office of Scientific Research (AFOSR) via IOE Grant No.~FA9550-19-1-7044 LASCEM, D.V.V by a joint-project grant from the FWF (Grant No. I04426, RSF/Russia 2019), R.v.B and P.Z. by the European Union’s Horizon 2020 research and innovation programme under Grant Agreement No. 817482 (PASQuanS), and R.v.B by the Austrian Research Promotion Agency (FFG) contract 884471 (ELQO). P.Z. acknowledges funding by the the European Union’s Horizon 2020 research and innovation programme under Grant Agreement No. 731473 (QuantERA via QTFLAG), and by the Simons Collaboration on Ultra-Quantum Matter, which is a grant from the Simons Foundation (651440). Innsbruck theory is a member of the NSF Quantum Leap Challenge Institute Q-Sense. The computational results presented here have been achieved (in part) using the LEO HPC infrastructure of the University of Innsbruck.

All statements of fact, opinions or conclusions contained herein are those of the authors and should not be construed as representing the official views or policies of the funding agencies.

\subsection*{Author contributions}
Ch.D.M. lead writing of the manuscript with assistance from R.K, D.V.V., R.v.B., and P.Z., and input from all co-authors. Ch.D.M., T.F., and I.P. built the experiment. Ch.D.M. and T.F. performed measurements. R.K., D.V.V., and P.Z. conceived of the method and provided theory. R.K. and R.v.B. developed the optimizer routines and implementation. Ch.D.M. and R.K. analysed the data. P.S., R.B., and T.M. supervised the experiment.

\subsection*{Competing interests}
The authors declare no competing interests.

\subsection*{Supplementary Material}
Supplementary Information is available for this paper.

\end{document}


\title{Supplementary Material:\\ Optimal metrology with programmable quantum sensors}
\author{Christian D. Marciniak$^{1}$}
\thanks{These authors contributed equally to this work}
\author{Thomas Feldker$^{1}$}
\thanks{These authors contributed equally to this work}
\author{Ivan Pogorelov$^1$}
\author{Raphael Kaubr{\"u}gger$^{2,3}$}
\author{Denis V. Vasilyev$^{2,3}$}
\author{Rick van Bijnen$^{2,3}$}
\author{Philipp Schindler$^1$}
\author{Peter Zoller$^{2,3}$}
\author{Rainer Blatt$^{1, 2}$}
\author{Thomas Monz$^{1, 4}$}
\email{thomas.monz@uibk.ac.at}
\affiliation{$^1$ Institut f{\"u}r Experimentalphysik, 6020 Innsbruck, Austria}
\affiliation{$^2$ Institute for Quantum Optics and Quantum Information, 6020 Innsbruck, Austria}
\affiliation{$^3$ Center for Quantum Physics, 6020 Innsbruck, Austria}
\affiliation{$^4$ Alpine Quantum Technologies (AQT), 6020 Innsbruck, Austria}

\maketitle

\section*{Supplementary Methods}
\subsection{State preparation and readout}
In the AQTION platform  \cai{} ions are hosted in a Paul trap forming a linear ion crystal under appropriate cooling. Optical qubits are encoded in the ground state $\ket{\textrm{4\,S}_{1/2},m_J = -1/2}$ and excited state $\ket{\textrm{3\,D}_{1/2},m_J = -1/2}$, which are connected via an electric quadrupole transition near \SI{729}{\nano\meter}.

Further lasers at \SI{397}{\nano\meter}, \SI{854}{\nano\meter}, and \SI{866}{\nano\meter} are required for state preparation and cooling. All ions are prepared in the qubit ground state via optical pumping before the start of each Ramsey sequence. The chain is cooled close to the motional ground state using resolved sideband cooling. Readout of the qubit system is performed optically via state-selective fluorescence near \SI{397}{\nano\meter}, rendering the ground state bright and excited state dark. We perform site-selective readout of each ion using spatially-resolved detection on a camera through a high-numerical-aperture objective solely to increase detection fidelity; the scheme itself requires only collective measurements. All qubit operations as part of the Ramsey sequence are performed via a single laser beam oriented along the ion chain. The beam is aligned to interact identically with all ions. 


\subsection{Implementation and calibration of unitaries}
Qubit rotations $\mathcal{R}_{x, y}$ are implemented via resonant excitation on the qubit transition. The relative phase of pulses used to implement driven excitations sets the basis in which the rotation takes place, where a phase shift of $\pi/2$ changes between $x$ and $y$. \Rz{} rotations may be implemented using AC Stark shifts induced by off-resonant excitation, but are not required here.

The infinite-range one axis twistings are implemented using the \MS{} interaction, whose unitary propagator takes the form $\exp{(-i\chi \hat{J}_x^2)}$ once appropriate conditions are met~\cite{roos2008ion}. The interaction is generated by a bichromatic light field detuned from the ion chain's first axial center-of-mass vibrational sidebands by $\Delta_\text{MS}$. An optical phase shift is again used to change between the $x$ and $y$ basis. Negative rotation angles may be implemented by inverting both tones' detunings with respect to the chosen mode's sidebands, but we restrict ourselves to positive twisting angles for ease of implementation.

The twisting angles implemented by the \MS{} interactions are given by the geometric phase $\chi$ enclosed during the operation. It is challenging to optimize this phase to an arbitrary value given that it has no easily accessible observable associated with it. Additionally, we require that the \MS{} operation leave no residual spin-motional entanglement which puts further conditions on experimental parameters that achieve any given value of $\chi$. Consequently, our strategy has two steps. First, we optimize a fully entangling \MS{} gate which implements a geometric phase of $\chi=\pi/2$ and creates a GHZ state. The state fidelity of the GHZ state allows for careful calibration of $\chi$ at this value. After this, we scale all parameters relative to the obtained values as required to maintain loop closure to the desired geometric phases.

In particular, we close eight loops in phase space after the $\tau_\text{MS} = \SI{1600}{\micro\second}$ gate duration. A single loop in \SI{200}{\micro\second} \textit{at the same detuning} thus implements a twisting angle of $\pi/16$. All required twisting angles are then obtained from this calibration by scaling the gate duration and detuning \textit{together} fixing $\tau_\text{MS} = 2\pi/\Delta_\text{MS}$, noting that then $\chi\propto\tau_\text{ms}^2$. The laser-induced AC Stark shifts during the \MS{} interaction are mitigated using power imbalancing~\cite{kirchmair2009deterministic} rather than centerline detuning. This dramatically reduces the complexity of the sequences since these phase shifts depend on the twisting angles, which are varied across the sequences, thus requiring individual measurement and compensation.

The proposal's original unitaries contain twisting operations partly along the $z$ axis. Unfortunately, \Tz{} cannot be natively implemented using the \MS{} interaction. However, we may use a simple basis change on the entire sequence to transform into solely using \Tx{} and \Ty{}. The Ramsey sequence unitary then takes the form
\begin{equation}
    \mathcal{U}_\text{Ramsey}' = \mathcal{U}_\text{De}^{x,y}\Ry{\phi}\mathcal{U}_\text{En}^{x,y}\Ry{\pi/2},
    \label{eq:Sequence}
\end{equation}
where the superscripts indicate the two bases used for twisting and rotation. In atomic clock settings the phase $\phi$ accumulated due to a detuning from the atomic reference frequency leads to a rotation $\Rz{\phi}$, which may be implemented using the identity $\Ry{\phi} = \Rx{-\pi/2}\Rz{\phi}\Rx{\pi/2}$.


\subsection{Restrictions on sequence parameters}
\label{sec:Restrictions}
It is difficult to carefully calibrate twisting angles that differ strongly in magnitude. Consequently, we restrict the optimizer search to twisting angles $\varphi_\text{min} \leq \varphi \leq \pi/8$. For small particle numbers, this comes at little cost to the predicted BMSE. $\varphi_\text{min}$ in turn is a lower bound on the twisting angles that is set by the minimal duration a \MS{} pulse can have in the experiment. This limitation is not set by experimental hardware, but instead by the pulse shaping necessary to minimize off-resonant carrier excitations~\cite{kirchmair2009deterministic}. $\varphi_\text{min}$ depends on the axial trap frequency, and we set it conservatively to $\varphi_\text{min} = \pi/160$. The absolute limit is near $\varphi_\text{min} = \pi/576$ at which point the pulse will be entirely made up of shaping slopes. As an example, this $\pi/160$ limit means that 2 of the 6 twistings in the $(1, 2)$ sequence are dropped for up to 26 ions, with changes in the theoretical BMSE well below the experimental uncertainty. For the $(1, 0)$ with up to 26 ions 1 of the 2 twistings are dropped, making this identical to a spin-squeezed state.

Rotations are not in principle limited in this fashion, but due to technical constraints in the hardware must also have a minimum duration. We mitigate this by calibrating multiple rotation objects with different Rabi rates for large and small rotations. Rotation angles below a certain threshold will nonetheless be skipped, adding experimental error for small rotation angles.

Adding additional entangling and decoding layers to a sequence increases the prospective gain relative to the CSS baseline Ramsey sequence. However, these gains diminish for ever larger layer depths at a fixed particle number as the sequence converges towards the optimal quantum interferometer, while both sequence overhead and gate error accumulation do not saturate in this manner. Consequently, there is a break-even point after which adding additional layers will decrease performance. For the particle numbers considered here this is at the (1, 2) sequence, as seen in Fig.~2 \textbf{c} and its inset, where the gain in adding an additional layer is already below the deviation of the data from the theory.


\subsection{Experimental calculation of cost}
The measurement record from the experiment is given by binary bit strings representing the $z$ projections of the ion chain state during the measurement. An example would be a string $0010 = \textrm{DDSD}$ representing a four ion register where ions 1, 2, and 4 are in the excited $\textrm{D}$ state and thus do not fluoresce, and ion 3 is in the bright ground state $\textrm{S}$.

For the measurement protocol we only require knowledge of collective measurement outcomes. Consequently, we calculate the probability $p(k|\phi_\text{exp})$ of finding $k$ excitations among all bit strings for a given phase $\phi_\text{exp}$ from the histogram of outcomes. This is identical to finding the probability  $p(m|\phi_\text{exp})$ to measure the excitation imbalance $m = (2k - N)/2$. Here $\phi_\text{exp}$ is the phase $\phi$ we want to implement up to a phase shift $\Tilde{\phi}$ that arises in the implementation due to experimental imperfections such as timing jitter or residual AC Stark shifts.

To calculate the MSE and BMSE we require the phase estimator $\phi_\text{est} = a m$, and knowledge of $\phi = \phi_\text{exp} - \Tilde{\phi}$ to obtain the conditional probabilities at the true phase $\phi$. To do this, we use the raw measurement data to calculate the conditional excitation imbalance probabilities and the MSE with both slope $a$ and $\tilde{\phi}$ as free parameters, at a fixed grid of phase values. The parameter-dependent MSE errors are integrated using the Simpson rule and the resulting BMSE is minimized in terms of the free parameters.
Typically we find \mbox{$\Tilde{\phi}\lesssim 0.01$}, and the slope $a$ is close to the theoretically predicted value. As an example, SI Tab.~\ref{tab:slopes} shows $1/a$ for the $\delta\phi = 1$ data point in Fig.~2 \textbf{c}.

\begin{table}[h]
    \centering
    \begin{tabular}{c|c|c}
        $(n_\text{En}, n_\text{De}$) & Exp. $1/a$ & Theory $1/a$  \\\hline
         $(0, 0)$ & \SI{4.731(12)}{} & \SI{4.745}{} \\
         $(1, 0)$ & \SI{4.363(9)}{} & \SI{4.378}{} \\
         $(0, 2)$ & \SI{3.042(4)}{} & \SI{3.001}{} \\
         $(1, 2)$ & \SI{3.023(4)}{} & \SI{2.963}{} \\
    \end{tabular}
    \caption{Comparison of linear phase estimator slopes $a$ determined experimentally via numerical optimization of raw data and from theory for $\delta\phi = 1$ data in Fig.~2 \textbf{c}.}
    \label{tab:slopes}
\end{table}

The cost function evaluation during the optimization is accelerated by measuring at $\phi$ values chosen to coincide with the nodes of the Hermite-Gaussian polynomials up to twelfth order. We further leverage the symmetry of the MSE around $\phi=0$ to perform Hermite-Gauss quadrature (HGQ) integration for $\phi\geq0$ only. The HGQ proves particularly useful if only a small number of $\phi$ points is used for the numerical integration, where equidistantly spaced integration schemes can lead to a biased estimate of the cost function. 

\subsection{Optimizer and meta-model design and tuning}

On-device optimization is essentially a classical optimization problem, with data generated on a quantum machine. This is a well-studied problem for which several solutions have been demonstrated~\cite{Cerezo2020}. For the optimization we employ a modified version of the Dividing Rectangles algorithm (DIRECT) \cite{Jones1993, nicholas2014DIRECT, Kokail2019} which is a gradient-free algorithm that guarantees global convergence for sufficiently many iterations. DIRECT divides the search space into hyperrectangles, which we will refer to as 'cells'. Each cell is represented by a single cost function evaluation taken in its interior. Promising cells are sampled at a finer scale by subdividing further, prioritizing cells with low cost function values, as well as cells with large size. 

During the optimization the algorithm maintains an internal representation, denoted meta-model, of the cost function landscape in the form of a Gaussian process \cite{Rasmussen2004}, modelling the data as jointly-distributed Gaussian variables with a trigonometric covariance kernel. For this meta-model we may expand the cost for each gate in a series of trigonometric functions of the interaction parameter since all interactions in the sequence are global. The cost function is then a product of $3(n_\text{En} + n_\text{De})$ trigonometric series which generally have non-vanishing expansion coefficients and argument scalings for both even and odd terms. The number of terms in each series as well as the scalings depend on the differences of eigenvalues of the generators of the gate, that is the differences of eigenvalues of $J_{x,y,z}$, and $J_{x,y,z}^2$. For rotations, these differences grow linearly in number, while for twistings they grow quadratically. The number of terms and scalings are known a priori for a given number of particles, so the cost has $6(n_\text{En} + n_\text{De})$ free parameters whose optimal match to the experimental data is continuously adapted with incoming measurement data.

The algorithm is provided with a finite measurement budget that it can spend during the optimization process. We invest only a relatively low number of measurements at each point to use the budget economically. If the variance of the cost function measurements is above a certain threshold, the algorithm can request refinement steps at points already sampled in order to increase the probability of correctly deciding which cells to subdivide. The algorithm selects when to perform the refinement steps, and how many measurements to spend in this stage. These decisions are based on methods from decision theory and optimal computing budget allocation~\cite{nicholas2014DIRECT, OCBA2008}.

To start the optimization we initialize search boxes ranging from \SI{50}{\percent} to \SI{150}{\percent} of the theoretically optimal parameter values. This situation is representative if there is some degree of trust in the operations implemented on the quantum machine. We further expect that for classically not simulatable resources, like finite-range interactions, the optimal parameters will smoothly change with the particle number, allowing extrapolation of similar constraints for the search space. We initialize the search by adding a random displacement to the search box relative to the theoretical optimum to prevent the optimization algorithm from trivially finding the theoretically optimal parameters by initially sampling the parameter space in the center of the search box. An example set of optimal parameters after a complete search together with theory predictions is given in SI Tab.~\ref{tab:parameters}.

\begin{table*}
    \renewcommand\arraystretch{1.1}
    \centering
    \begin{tabular}{c || c | c c c | c c c | c c c}
         &  & \multicolumn{3}{c|}{Entangling} &  \multicolumn{3}{c|}{Decoding 1} & \multicolumn{3}{c}{Decoding 2}\\
        $(n_\text{En}, n_\text{De})$ & Approach & \Ty{} & \Tx{} & \Rx{} & \Rx{} & \Tx{} & \Ty{} & \Rx{} & \Tx{} & \Ty{}\\\hline\hline
        \multirow{2}{*}{$(1, 0)$} & Theory & 0.0551 & 0 & -1.0699 & 0 & 0 & 0 & 0 & 0 & 0\\
                                  & Optimizer & 0.1080 & 0 & -1.3090 & 0 & 0 & 0 & 0 & 0 & 0\\\hline
        \multirow{2}{*}{$(1, 2)$} & Theory & 0.0626 & 0 & -1.2725 & 0.6938 & 0.0196 & 0.0631 & -0.6518 & 0.0196 & 0\\
                                  & Optimizer & 0.0896 & 0 & -1.3818 & 0.5841 & 0.0213 & 0.0813 & -0.6357 & 0.0196 & 0\\\hline
    \end{tabular}
    \caption{Twisting and rotational angles (in rad) of variational unitaries for two variational sequence types with 26 particles. Optimizer parameters are for the data displayed in Fig.~3 in the main text. Twisting angles of 0.0196 correspond to the lower boundary of $\pi/160$ as described in section~\ref{sec:Restrictions}.}
    \label{tab:parameters}
\end{table*}


\subsection{Coarse and fine optimizer scans}
Any optimizer intrinsically requires many evaluations of the same parameter landscape over time. In real experiments the quantum operations are subject to drift. Consequently, we perform `coarse' evaluations to speed up the procedure with fewer repetitions (typically 100) per point and fewer sampling points in $\phi$ per evaluation (typically 5 - 11 nodes). Once a good parameter candidate has been identified, the optimizer requests a `fine' scan of these parameters with finer sampling (typically 20 or 21 nodes) on both sides of $\phi = 0$, and more repetitions (typically 250).

Longer ion chains are subject to more frequent collisions with background gas in the trap than smaller chains, leading to frequent melting of the crystal. A molten crystal has to be recrystallized before the next state initialization can begin - a process typically referred to as refreezing. Refreezing the crystal takes substantially longer than state preparation and sequence execution. Refreeze events requiring remeasuring of a given point are more frequent in long chains, and fine scans with many repetitions per measurement point on long chains can thus become time consuming. Additionally, a detected refreeze event will discard all measured repetitions and start the measurement process anew. We alleviate this problem by batching measurement requests from the optimizer into multiples of 50-repetition measurements, so the maximum number of discarded measurements is 50. 

\subsection{Noise frequency reconstruction}
We try to reconstruct the known value of injected noise (detuning) by performing phase estimation with two different Ramsey sequences, while also accounting for the drift in the laser's reference cavity.

A frequency detuning $\Delta\omega$ is drawn at random from a normal distribution with width $\Delta\omega_N = 2\pi\cdot\SI{40}{\hertz}$ and added to the laser frequency using an acousto-optic modulator. The distribution is truncated at $2\Delta\omega_N$ to reduce the influence of rare large detunings on the measurement outcome from the limited statistics of 200 different samples per Ramsey time.

We perform three Ramsey sequences successively where we impart the phase $\phi$ via \mbox{\Rx{-\pi/2}\Rz{\phi=\Delta\omega T_R}\Rx{\pi/2}}, with a detuning $\Delta\omega$: First, a standard Ramsey sequence without detuning ($\Delta\omega=0$), with $T_R = \SI{15}{\milli\second}$ and averaging over 50 repetitions. This is used to correct for the slow drift of our qubit laser's cavity. 
Second, 50 measurements with detuning $\Delta\omega$ using a standard CSS Ramsey sequence where the Ramsey time is varied and the laser frequency is estimated from evaluation of single measurements. Lastly, a sequence optimized for a prior width $\delta \phi = 0.6893$ with depths $(1, 2)$ is used analogously to the second CSS sequence. This value is chosen as the equivalent Ramsey time at which both the CSS and the optimized sequence approximately have their respective minima in the BMSE. We repeated this measurement 200 times for each Ramsey time. The order of CSS and $(1, 2)$ sequence was reversed after 100 repetitions to exclude a systematic bias, but no difference in outcome was found.  

To explain the difference between the ideal measurement and our experiment we identify three main sources of error: First, frequency flicker noise of our laser. To determine the frequency noise we perform the measurement described above, without detuning $\Delta\omega$. For long Ramsey times the standard deviation of the measured frequencies converges to $b_\alpha\approx2\pi\cdot\SI{6}{\hertz}$. 
Second, the non-ideal BMSE measured in our system. 
Third, the BMSE increases with detuning $\Delta\omega$, so that the distribution width $\Delta\omega_N$ is a compromise between reducing the influence of laser frequency noise on the measurement ($\Delta\omega \gg b_\alpha$) and avoiding gate errors to achieve the best BMSE ($\Delta\omega \rightarrow 0$). Consequently the achievable gain is directly correlated to the laser frequency noise.

We note that the deterioration of the BMSE with detuning is an artifact of our measurement. In an optical clock the detuning $\Delta\omega$ is small and $T_R$ is consequently much larger to reach the same $\delta\phi$. The large values of $\Delta\omega$ here cause deterioration of the gate-based implementation of the twistings, which is a consequence of the technical limitations of the implementation, not the scheme itself.

\subsection{Sequence overhead}
The variational sequences contain more operations than the archetypical Ramsey sequence, which produces time overhead in state preparation and measurement relative to the coherent spin state interferometer. These effects are typically negligible for high-precision trapped-ion or atomic systems, in particular for atomic clocks, where Ramsey times of order \SI{1}{\second} are reached. Even in our case the increase in sequence length accounts for less than \SI{10}{\percent} of the state preparation time, below the gain in $T_R$. Specifically, for the sequences (1, 0) (SSS) our overhead is \SI{165}{\micro\s}, the (1, 2) takes an additional \SI{250}{\micro\s}, and (2, 1) (GHZ) takes \SI{300}{\micro\s} extra compared to the CSS.

\section*{Supplementary Discussion}

\subsection{Comparison of phase estimator functions}

\begin{figure}[ht]
    \centering
    \includegraphics[width=\columnwidth]{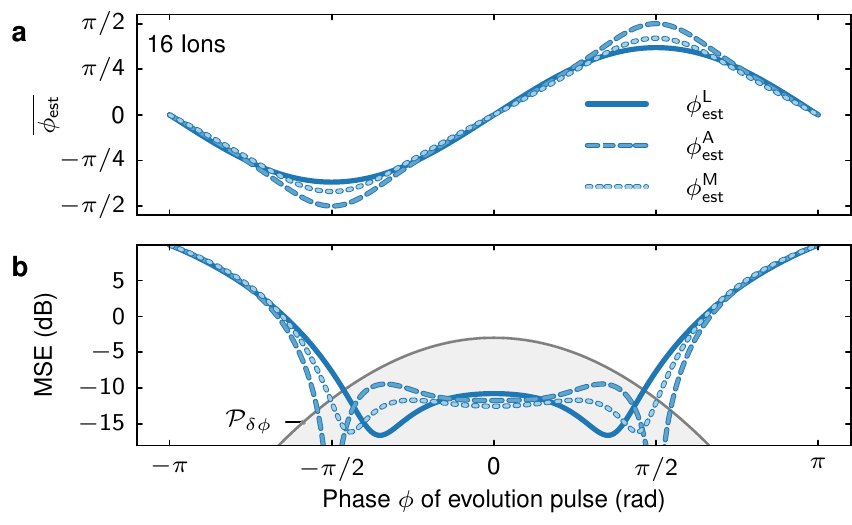}
    \caption{Theoretical performance of a conventional Ramsey sequence $(n_{\rm En}, n_{\rm De}) = (0,0)$ for different estimator functions. \textbf{a} Expectation value $\overline{\phi_{\rm est}}$ of phase estimators as a function of evaluation pulse phase $\phi$. Any deviation from a line with unit slope is a manifestation of bias. \textbf{b} MSE calculated for the different estimator functions. Overlaid in grey is the prior distribution $\mathcal{P}_{\delta\phi}\left(\phi\right)$ with $\delta\phi \approx 0.79$. $\phi^{\rm L}$ and $\phi^{\rm M}$ are optimized for this particular prior distribution, while $\phi^{\rm A}$ is independent of it. $\Delta\phi/\delta\phi$ values for linear, arcsine, and MBMSE corresponding the MSE and $\mathcal{P}_{\delta \phi}$ in \textbf{b} are: \SI{-4.01}{\dB}, \SI{-3.94}{\dB}, \SI{-4.20}{\dB}.}
    \label{fig:ESTMSE}
\end{figure}

\begin{figure}[h!]
    \centering
    \includegraphics[width=\columnwidth]{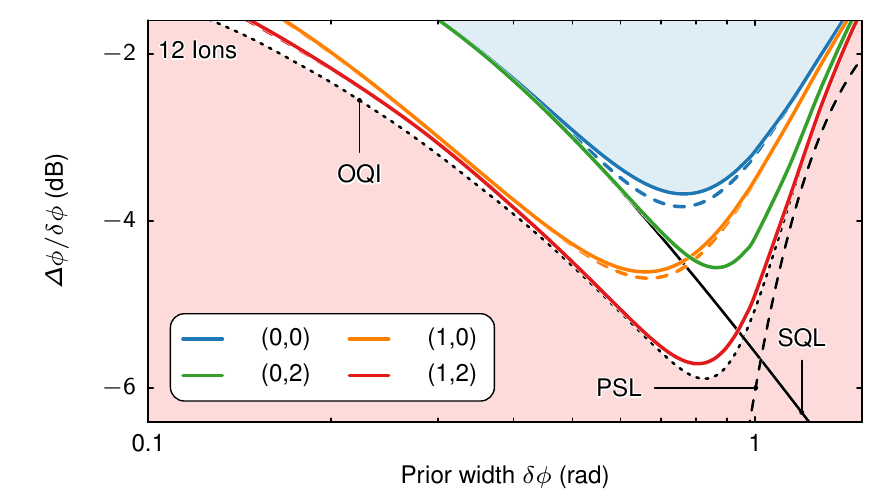}
    \caption{Theoretical $\Delta\phi/\delta\phi$ as a function of prior width $\delta\phi$ for 12 ions in the four variational sequences. The blue shaded region corresponds roughly to classical Ramsey sequences. The red shaded region is inaccessible using this scheme, where the boundary corresponds to the optimal quantum interferometer (OQI). The OQI, the standard quantum limit (SQL) and the phase slip limit (PSL) are indicated by black lines. Solid lines are obtained for a linear phase estimator while the dashed lines are obtained for the estimator that minimizes the $\Delta\phi$ over all possible estimators at the respective prior width.}
    \label{fig:BMSE}
\end{figure}

Estimator functions are used to produce an estimate of the desired parameter based on measurement outcomes. Choosing an appropriate estimator function is paramount in metrology. This choice is informed not only by suitability in the ideal case, but also implementation considerations such as resilience against noise, or practicability of computation in systems with time constraints.

In this study we judge the performance of different types of variational sequences relative to one another based on our cost function which in turn depends on the choice of estimator function. In the main text we restrict ourselves to considering a linear estimator of the phase $\phi$ of the form
\begin{equation}
    \phi^{\rm L}_{\rm est}(m) = a\,m.
\end{equation}
However, other choices of estimator function are possible, and may at first glance seem superior. For sinusoidally varying measurement outcomes one may consider using the analytic inverting function, that is using an arcsine estimator 
\begin{equation}
    \phi^{\rm A}_{\rm est}(m) = \arcsin(m \tfrac{2}{N}), 
\end{equation}
which ideally is an unbiased estimator for all $\phi$ for sufficiently large particle number $N$. It is clear that any loss of amplitude through e.g. decoherence effects will cause bias everywhere again, i.e. this estimator feature is not stable in the presence of noise.

Another choice is the minimum Bayesian mean squared error (MBMSE) estimator given by
\begin{equation}
    \phi^{\rm M}_{\rm est}(m) = \int_{-\infty}^{\infty} d\phi\, \phi\, p(\phi|m),
\end{equation}
where $p(\phi|m)  = p(m|\phi)  \mathcal{P}_{\delta \phi}(\phi) / p(m)$ may be obtained from measured conditional probabilities $p(m|\phi)$ using Bayes' theorem, and with $p(m) = \int_{-\infty}^{\infty} d\phi\, p(m|\phi)$.
This estimator can be shown to minimize the cost function, i.e. the BMSE, over all possible estimators~\cite{Demkowicz2015}, but is computationally expensive.

Both the arcsine and MBMSE estimators extend the dynamic range of an an ideal, conventional $(0, 0)$ Ramsey interferometer relative to the linear estimator, that is they remain approximately unbiased for a larger range of parameter values. At 16 ions the MBMSE extends only slightly further than the linear estimator, while the arcsine extends the dynamic range to almost the interval $(-\pi/2, \pi/2)$ (SI Fig.~\ref{fig:ESTMSE} \textbf{a}). This ordering of dynamic ranges is preserved as the particle number is increased. 
However, the MSE as a metric to judge estimator performance includes not only the bias but also the sensitivity of the interferometer. Consequently, when $\phi=\pm\pi/2$ and the interferometer is in an $z$ eigenstate, projection noise vanishes and at zero bias the arcsine estimator MSE vanishes as well, while the linear and MBMSE estimators remain finite (SI Fig.~\ref{fig:ESTMSE} \textbf{b}). While in the immediate vicinity of $\phi=0$ the arcsine estimator performs better than the linear, its MSE remains at a similar level further out. The sensitivity of the linear estimator on the other hand is generally higher in the region where the prior distribution is appreciable.  As a result, the arcsine estimator performs overall worse than both linear and MBMSE estimators in our operational cost function $\mathcal{C}$, which averages over the MSE weighted by the prior distribution $\mathcal{P}_{\delta \phi}$. 

Returning to the normalized cost $\Delta\phi/\delta\phi$ as a metric we see that the MBMSE estimator retains a small advantage over the simple linear estimator at prior widths close to the sequences' optimal widths (SI Fig.~\ref{fig:BMSE}). This gain decays quickly for prior widths away from the optima, and is completely lost for sequences that feature decoding layers, i.e. $n_\text{De}>0$. We can thus conclude that concentrating on the simple and stable linear estimators does not qualitatively change the analysis, and is near-optimal for the more complicated sequences.


\subsection{Application to atomic clocks}
A primary example where programmable quantum sensors may directly benefit a crucial application in the near-term is the improvement of atomic or ionic clocks. The resources and interactions available in for example optical lattice clocks have already been studied theoretically~\cite{kaubruegger2019variational, theory_posse}, and their application demonstrated in the context of squeezing for enhanced clock operation~\cite{pedrozo2020entanglement}. Programmable quantum sensors offer additional benefits to atomic clocks in realistic metrology settings. The tailored entanglement extends the interferometer dynamic range, for example, which increases the optimal Ramsey time $T_R$ at a fixed noise level. Longer Ramsey times can further increase the duty cycle of atomic clocks which reduces the optical Dick effect~\cite{takamoto2011frequency, schioppo2017ultrastable}. In SI Tab.~\ref{tab:AllanGains} we show the potential gain of the scheme for 12 and 26 ions as demonstrated here compared to 362 atoms as in Ref.~\cite{pedrozo2020entanglement} in an optical lattice.

\begin{table}[h]
    \renewcommand\arraystretch{1.1}
    \centering
    \begin{tabular}{c || c | c | c || c}
        $N$ & Approach  & $(1, 0)$ & $(1, 2)$ & Gap to OQC\\\hline\hline
        \multirow{2}{*}{12} & Theory & \SI{1.49(0)}{\dB} & \SI{2.13(0)}{\dB} & \SI{0.04(0)}{\dB}\\
        & Direct & \SI{1.38(1)}{\dB} & \SI{1.75(2)}{\dB} & \SI{0.50(5)}{\dB}\\\hline
        \multirow{3}{*}{26} & Theory & \SI{2.12(0)}{\dB} & \SI{2.70(0)}{\dB} & \SI{0.80(0)}{\dB}\\
        & Direct & \SI{1.47(8)}{\dB} & \SI{2.02(8)}{\dB} & \SI{1.66(8)}{\dB}\\
        & Optimizer & \SI{1.54(9)}{\dB} & \SI{1.77(8)}{\dB} & \SI{1.90(8)}{\dB}\\\hline
        362 & Theory & \SI{4.53(0)}{\dB} & \SI{7.50(0)}{\dB} & \SI{1.47(0)}{\dB}
    \end{tabular}
    \caption{Gain over CSS in Allan deviation for 12 and 26 ions (theoretical and closest-measured value of $\delta\phi$) and 362 atoms (theoretical) for two different interferometer states at their respective optimal prior widths (Ramsey times). Gap of $(1, 2)$ sequence to optimal quantum clock (OQC) shown alongside. Deeper circuits with more layers approach the OQC more closely.}
    \label{tab:AllanGains}
\end{table}

The potential metrological gain in the clock setup can be increased by \SI{3}{\dB} by using the slight additional resource overhead of the decoding layers, but notably using the same operations as before. Similarly, Ref.~\cite{pedrozo2020entanglement} points out the first squeezing demonstrated on an optical transition with a Wineland squeezing parameter of \SI{4.4(6)}{\dB} with $\approx360$ atoms. We directly measure a Wineland squeezing parameter~\cite{wineland1994squeezed} of \SI{5.29(5)}{\dB} below the standard quantum limit using only 26 particles with the $(1,2)$ sequence.  Second, the robustness of the scheme with respect to variations of the prior width (Fig.~4) enables reliable operation when shot-per-shot measurements exhibit fluctuations in the number of particles as commonly encountered in cold atom experiments.